\documentclass[12pt,preprint]{aastex}
\usepackage{epsf}

\def\linebreak{\hfil\break}

\def\
{\hfil\linebreak}

\def\ang{\ifmmode {\rm \mathring{A}} \else {$\rm \mathring{A}$}\fi}
\def\deg{\ifmmode {^\circ}\else {$^\circ$}\fi}
\def\degree{\ifmmode {^\circ}\else {$^\circ$}\fi}
\def\mum{\ifmmode {\rm \mu {\rm m}}\else $\rm \mu {\rm m}$\fi}
\def\arcsec{\ifmmode ^{\prime \prime}\else $^{\prime \prime}$\fi}

\def\inch{\ifmmode ^{\prime \prime}\else $^{\prime \prime}$\fi}
\def\arcmin{\ifmmode ^{\prime}\else $^{\prime}$\fi}
\def\qprime{\ifmmode q^{\prime}\else $q^{\prime}$\fi}

\def\degree{\ifmmode {^\circ}\else {$^\circ$}\fi}
\def\arcsec{\ifmmode ^{\prime \prime}\else $^{\prime \prime}$\fi}

\def\inch{\ifmmode ^{\prime \prime}\else $^{\prime \prime}$\fi}
\def\arcmin{\ifmmode ^{\prime}\else $^{\prime}$\fi}

\def\mjup{\ifmmode { M_J}\else $ M_J$\fi}
\def\rjup{\ifmmode { R_J}\else $ R_J$\fi}
\def\mearth{\ifmmode { M_{\oplus}}\else $ M_{\oplus}$\fi}
\def\rearth{\ifmmode { R_{\oplus}}\else $ R_{\oplus}$\fi}
\def\ldust{\ifmmode { L_d}\else $ L_d$\fi}
\def\ldstar{\ifmmode { L_d / L_{\star}}\else $ L_d / L_{\star}$\fi}
\def\lstar{\ifmmode { L_{\star}}\else $ L_{\star}$\fi}
\def\lsun{\ifmmode { L_{\odot}}\else $ L_{\odot}$\fi}
\def\mlz{\ifmmode m_{l,0}\else $ m_{l,0}$\fi}
\def\bl{\ifmmode b_l\else $ b_l$\fi}
\def\mcl{\ifmmode M_c\else $ M_c$\fi}
\def\xcl{\ifmmode M_c/M_p\else $ M_c/M_p$\fi}
\def\mp{\ifmmode M_p\else $ M_p$\fi}
\def\mpl{\ifmmode M_p\else $ M_p$\fi}
\def\mstar{\ifmmode M_{\star}\else $ M_{\star}$\fi}
\def\msun{\ifmmode M_{\odot}\else $ M_{\odot}$\fi}
\def\tstar{\ifmmode T_{\star}\else $ T_{\star}$\fi}
\def\rstar{\ifmmode R_{\star}\else $ R_{\star}$\fi}
\def\rsun{\ifmmode R_{\odot}\else $ R_{\odot}$\fi}
\def\mjup{\ifmmode M_{J}\else $ M_{J}$\fi}
\def\rjup{\ifmmode R_{J}\else $ R_{J}$\fi}
\def\mjupyr{\ifmmode { M_J~yr^{-1}}\else $ M_J~yr^{-1}$\fi}
\def\msunyr{\ifmmode { {\rm M_{\odot}~yr^{-1}}}\else $ {\rm M_{\odot}~yr^{-1}}$\fi}
\def\gyr{\ifmmode {\rm g~yr^{-1}}\else $\rm g~yr^{-1}$\fi}
\def\ergg{\ifmmode {\rm erg~g^{-1}}\else $\rm erg~g^{-1}$\fi}
\def\kms{\ifmmode {\rm km~s^{-1}}\else $\rm km~s^{-1}$\fi}
\def\ms{\ifmmode {\rm m~s^{-1}}\else $\rm m~s^{-1}$\fi}
\def\rhill{\ifmmode R_H\else $R_H$\fi}
\def\rfast{\ifmmode R_{fast}\else $R_{fast}$\fi}
\def\rgap{\ifmmode R_{gap}\else $R_{gap}$\fi}
\def\vhill{\ifmmode v_H\else $v_H$\fi}
\def\qdstar{\ifmmode Q_D^\star\else $Q_D^\star$\fi}
\def\mesc{\ifmmode m_{esc}\else $m_{esc}$\fi}
\def\rmin{\ifmmode r_{min}\else $r_{min}$\fi}
\def\rmax{\ifmmode r_{max}\else $r_{max}$\fi}
\def\mmax{\ifmmode m_{max}\else $m_{max}$\fi}
\def\rmind{\ifmmode r_{min,d}\else $r_{min,d}$\fi}
\def\rmaxd{\ifmmode r_{max,d}\else $r_{max,d}$\fi}
\def\mmaxd{\ifmmode m_{max,d}\else $m_{max,d}$\fi}
\def\qz{\ifmmode q_{0}\else $q_{0}$\fi}
\def\qi{\ifmmode q_{i}\else $q_{i}$\fi}
\def\ql{\ifmmode q_{l}\else $q_{l}$\fi}
\def\qs{\ifmmode q_{s}\else $q_{s}$\fi}
\def\r0{\ifmmode r_{0}\else $r_{0}$\fi}
\def\m0{\ifmmode m_{0}\else $m_{0}$\fi}
\def\M0{\ifmmode M_{0}\else $M_{0}$\fi}
\def\xm{\ifmmode x_{m}\else $x_{m}$\fi}
\def\gyr{\ifmmode {\rm g~yr^{-1}}\else ${\rm g~yr^{-1}}$\fi}
\def\cms{\ifmmode {\rm cm~s^{-1}}\else ${\rm cm~s^{-1}}$\fi}
\def\gcms{\ifmmode {\rm g~cm^{-2}}\else $\rm g~cm^{-2}$\fi}
\def\gcmc{\ifmmode {\rm g~cm^{-3}}\else $\rm g~cm^{-3}$\fi}

\def\2470{[24]--[70]}

\newbox\grsign \setbox\grsign=\hbox{$>$} \newdimen\grdimen \grdimen=\ht\grsign
\newbox\simlessbox \newbox\simgreatbox
\setbox\simgreatbox=\hbox{\raise.5ex\hbox{$>$}\llap
     {\lower.5ex\hbox{$\sim$}}}\ht1=\grdimen\dp1=0pt
\setbox\simlessbox=\hbox{\raise.5ex\hbox{$<$}\llap
     {\lower.5ex\hbox{$\sim$}}}\ht2=\grdimen\dp2=0pt

\begin{document}

\title{EG Andromedae: A New Orbit and Additional Evidence for a Photoionized Wind}
\vskip 7ex
\author{Scott J. Kenyon}
\affil{Smithsonian Astrophysical Observatory,
60 Garden Street, Cambridge, MA 02138} 
\email{e-mail: skenyon@cfa.harvard.edu}
\author{Michael R. Garcia}
\affil{NASA Headquarters, Mail Suite 3Y28,
300 E Street SW, Washington, DC 20546-0001}
\email{e-mail: michael.r.garcia@nasa.gov}
%
%-------------------------- ABSTRACT ----------------------------------
%
%\doublespace

\begin{abstract}
We analyze a roughly 20 yr set of spectroscopic observations for the symbiotic 
binary EG And. Radial velocities derived from echelle spectra are best-fit with
a circular orbit having orbital period $P$ = 483.3 $\pm$ 1.6~d and semi-amplitude 
$K$ = 7.34 $\pm$ 0.07~\kms. Combined with previous data, these observations rule 
out an elliptical orbit at the 10$\sigma$ level. Equivalent widths of H~I Balmer 
emission lines and various absorption features vary in phase with the orbital 
period. Relative to the radius of the red giant primary, the apparent size 
of the H~II region is consistent with a model where a hot secondary star with 
effective temperature $T_h \approx$ 75,000~K ionizes the wind from the red giant. 
\end{abstract}


\section{INTRODUCTION}

Symbiotic stars are interacting binary systems consisting of a 
red giant primary star, a hotter secondary star, and a surrounding 
ionized nebula \citep{allen1984,kenyon1986,belczy2000,corradi2008}. 
Typical systems have orbital periods ranging from $\sim$ 200~d 
to several decades, bright emission lines from H~I, He~I, He~II, 
C~IV, [Fe~VII], O~VI, and other highly ionized species, and 
occasional 2--3 mag eruptions.  The primary is usually on its 
first ascent of the red giant branch, but some of the primary 
stars are Mira variables.  Most secondary stars are hot white 
dwarfs powered by material accreted from the red giant wind
\citep[e.g.,][]{kw1984,murset1991,sokol2006}.  
In a few systems, an accreting main sequence \citep{kenyon1991}
or neutron \citep{chak1997,hinkle2006} star energizes the nebula.

EG And (HD 4174) is a low excitation symbiotic star, with bright 
H~I, [O~III], and [Ne~III] optical emission lines superimposed 
on an M-type absorption spectrum \citep{wilson1950,kenyon1986,
skopal1991,munari1993}.  In addition to a strong blue continuum, 
ultraviolet (UV) spectra show emission lines from He~II, C~IV, 
N~V, O~III, O~VI, and Si~IV \citep{stencel1984,oliver1985,vogel1991,
crowley2008}.  The He II $\lambda$1640 emission line flux and 
the UV continuum shape suggest a hot component with effective
temperature $T_h \, \sim \, 5-8 \, \times 10^4$~K 
\citep{kenyon1985,murset1991,kolb2004,crowley2008}. 
If the cool primary star is a normal M2~giant 
\citep{kg1983,oliver1985,kenyon1987a,kenyon1988a,kenyon1988b,fekel2000}, 
the mass of the secondary star is 
0.3--0.5~\msun\ \citep{vogel1991,vogel1992,kolb2004}.

Photometric and spectroscopic observations of EG~And yield an 
accurate orbital period of 482--483~d \citep{kaler1983,oliver1985,
munari1988,vogel1991,skopal1991,tomov1996,skopal1997,fekel2000,
skopal2007,jurdana2010,skopal2012}. 
At optical minimum, the red giant occults the hot secondary and the 
strong emission lines \citep{stencel1984,vogel1991,crowley2008}. 
Despite a high quality orbit from optical and infrared absorption 
lines \citep{fekel2000}, it is not clear whether the optical 
variability is due to ellipsoidal variations \citep{wilson1997}, 
illumination of the red giant photosphere or wind \citep{skopal2007,
skopal2008,skopal2012}, or colliding winds from the primary and 
secondary stars \citep{walder1995,tomov1995,calabro2014}.

Here we describe low and high resolution optical spectra of EG~And.
With roughly 20~yr of fairly continuous observations, our goal is
to refine the orbital parameters and to constrain models for time 
variations in the emission lines. After a short description of the data 
acquisition in \S\ref{sec: obs}, analysis of SAO radial velocities 
(\S\ref{sec: an-rv}) yields an orbital period of 483.3~$\pm$~1.3~d, 
with a semi-amplitude $K_1$ = 7.34~$\pm$~0.07~\kms\ for the red giant 
primary. Combined with previous optical and UV data, the new orbit
indicates the binary consists of a 0.35--0.55~\msun\ white dwarf and a 
1.1--2.4~\msun\ red giant at a distance of 400 pc (\S\ref{sec: an-pars}).
Equivalent widths of the H~I~Balmer lines and TiO absorption bands 
show clear evidence for illumination of the red giant wind by the 
hot secondary (\S\ref{sec: an-ew}--\ref{sec: an-wind}). Modest X-ray
fluxes suggest the hot white dwarf accretes a small fraction of the
red giant wind (\S\ref{sec: an-acc}).  We conclude with a brief summary 
(\S\ref{sec: disc}).

\section{OBSERVATIONS}
\label{sec: obs}

From 1994 September to 2016 January, P. Berlind, M. Calkins, 
and other observers acquired 480 low resolution optical spectra 
of EG And with FAST, a high throughput, slit spectrograph mounted 
at the Fred L. Whipple Observatory 1.5-m telescope on Mount Hopkins, 
Arizona \citep{fab1998}.  They used a 300 g mm$^{-1}$ grating blazed 
at 4750 \ang, a 3\arcsec~slit, and a thinned 512 $\times$ 2688 CCD.  
These spectra cover 3800--7500 \ang\ at a resolution of 6~\ang.  Spectra 
are reduced in NOAO IRAF\footnote{IRAF is distributed by 
the National Optical Astronomy Observatory, which is operated by the 
Association of Universities for Research in Astronomy, Inc. under contract 
to the National Science Foundation.}.  After trimming the CCD frames 
at each end of the slit, we correct for the bias level, flat-field 
each frame, and apply an illumination correction.  The full wavelength 
solution is derived from calibration lamps acquired immediately after 
each exposure.  The wavelength solution for each frame has a probable
error of $\lesssim$ $\pm$0.5~\ang.  To construct final 1-D spectra, 
object and sky spectra are extracted using the optimal extraction 
algorithm {\it apextract} within IRAF.  Most of the resulting spectra 
have moderate signal-to-noise, S/N $\gtrsim$ 15--30 per pixel.

On reasonably clear nights, observations of several standard stars enable
flux-calibration on the \citet{hayes1975} system \citep[see also][]{barnes1984,
massey1988}. After an extinction correction using the KPNO extinction curve, 
the sensitivity curve for each standard is derived using the appropriate 
tables in the IRAF irscal directory. After calculating the average sensitivity
curve for 6--10 standard stars, we apply the flux-calibration to spectra of
symbiotic stars.  On high quality nights, the calibration has a typical 
uncertainty of $\pm$0.10~mag ($\pm$0.05~mag) for $\lambda \lesssim$ 
4000--4500~\ang\ ($\lambda \gtrsim$ 4000--4500~\ang).

Despite the excellent quality of flux-calibration on some nights, many nights 
were plagued with cirrus or thicker clouds. To measure the long-term variation
of absorption and emission features, we measure equivalent widths using 
{\it sbands} within IRAF. In this routine, we define a central wavelength 
$\lambda_0$ for each feature and continuum points on either side of 
$\lambda_0$.  After measuring fluxes in 30~\ang\ bandpasses at all three
wavelengths and interpolating the continuum flux to the central wavelength, 
the equivalent width is
EW = $-2.5$~log~($F_0 / F_i$), where $F_i$ is the interpolated continuum flux 
and $F_0$ is the observed flux in a bandpass centered at $\lambda_0$. With
this definition, absorption (emission) lines yield a positive (negative)
equivalent width. 

For strong emission features, we derive fluxes using IRAF {\it splot}, which
performs least-squares fits of a gaussian plus a polynomial. Comparisons 
with fluxes derived from the equivalent widths suggest a 5\% to 10\% error.
As emission lines become weaker, identifying a robust continuum for the 
least-squares fit becomes more challenging.  The equivalent widths then 
appear to provide a more consistent measure of the strength of the 
absorption or emission in each line.

Prior to the start of the FAST observations, we obtained occasional optical 
spectrophotometric observations of EG And throughout 1982-1989 with the 
cooled dual-beam intensified Reticon scanner (IRS) mounted on the white 
spectrograph at the KPNO No. 1 and No. 2 90 cm telescopes.  We used NOAO 
IRAF software to reduce these data to the \citet{hayes1975} flux scale;
the photometric calibration has an accuracy of $\pm$3--5\%.

\begin{figure} 
\includegraphics[width=6.5in]{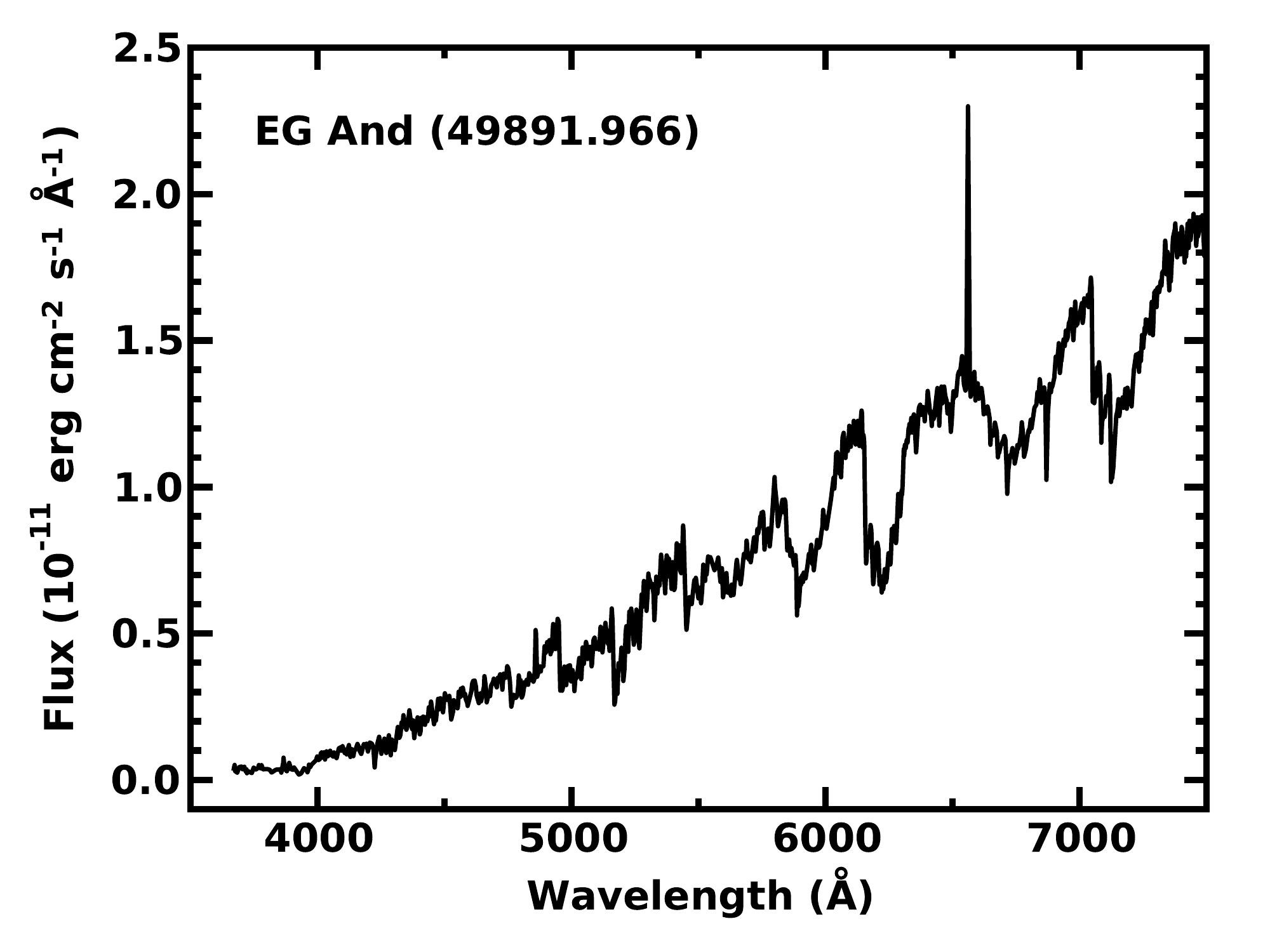}
\vskip 3ex
\caption{
Flux-calibrated optical spectrum of EG And acquired on
JD~2449891.966. Aside from strong TiO absorption bands,
the system has prominent H$\alpha$ emission and weak 
H$\beta$ and H$\gamma$ emission features. 
\label{fig: fast-spec}
}
\end{figure}

Fig.~\ref{fig: fast-spec} shows a flux-calibrated FAST spectrum for EG~And.
Aside from extending over a somewhat larger wavelength range, the IRS 
spectra are identical.  Prominent TiO absorption features are visible 
longwards of $\sim$ 5000 \ang; the Ca I $\lambda$4227 line is also very 
strong.  Low ionization emission lines are visible on many spectra. Aside
from H~I Balmer lines, weak Fe~II, [Fe~II], He~I, [O~III], and [Ne~III] 
are sometimes visible. Higher ionization features often observed in 
symbiotic stars, such as He~II, N~III, and [Fe~VII], are never visible
on FAST or IRS spectra.

Various remote observers acquired high resolution spectroscopic observations 
of EG~And with the echelle spectrographs and Reticon detectors on the 1.5-m 
telescopes of the Fred L. Whipple Observatory on Mount Hopkins, Arizona and 
the Oak Ridge Observatory in Harvard, Massachusetts \citep{latham1985}.
These spectra cover a 44 $\rm \ang$ bandpass centered near 5190~$\rm \ang$
or 5200~$\rm \ang$ and have a resolution of roughly 12 $\kms$.
\citet{garcia1986}, \citet{kg1986}, and \citet{garcia1988} discuss other 
details regarding the acquisition and reduction of these data for symbiotic 
stars. 

\begin{figure} 
\includegraphics[width=6.5in]{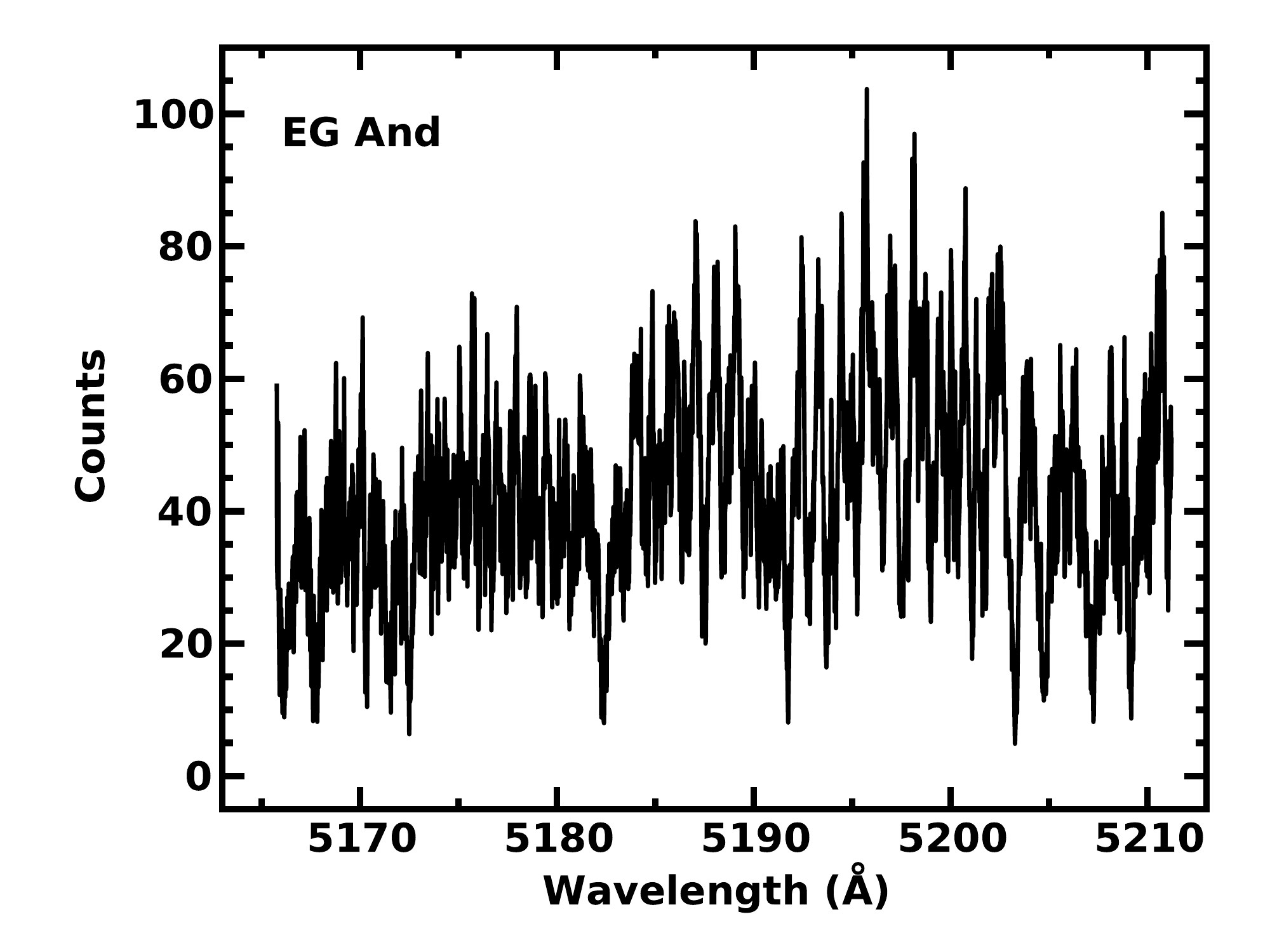}
\vskip 3ex
\caption{
Echelle spectrum of EG And. The strong absorption lines
are characteristic of early type M giant stars.
\label{fig: ech-spec}
}
\end{figure}

Fig.~\ref{fig: ech-spec} shows a typical echelle spectrum. The Mg~I~b
and other absorption lines characteristic of M giants are prominent.
Although some symbiotic stars have weak iron emission lines at these 
wavelengths, our EG And spectra never show these features 
\citep[see also][]{crowley2008}.  

We measure absorption-line radial velocities of the M giant component by 
cross-correlating EG And spectra against the spectrum of a template star
\citep{tonry1979,hartmann1986}.  All of the EG And spectra were 
cross-correlated against a very well-exposed spectrum of an M-type giant 
with velocity derived from cross-correlation against various IAU standard 
stars and the dusk/dawn sky \citep[e.g.,][]{latham1985,mazeh1996}.
This procedure places radial velocities from the two observatories on a 
common scale with typical errors of 0.75~\kms. After eliminating several 
spectra compromised by moonlight and cirrus and updating several 
measurements from \citet{garcia1986}, we have 108 velocities over roughly 
14 years of observations (Table~\ref{tab: rvs}).

\section{ANALYSIS}
\label{sec: an}

\subsection{Orbital Solution}
\label{sec: an-rv}

\begin{figure} 
\includegraphics[width=6.5in]{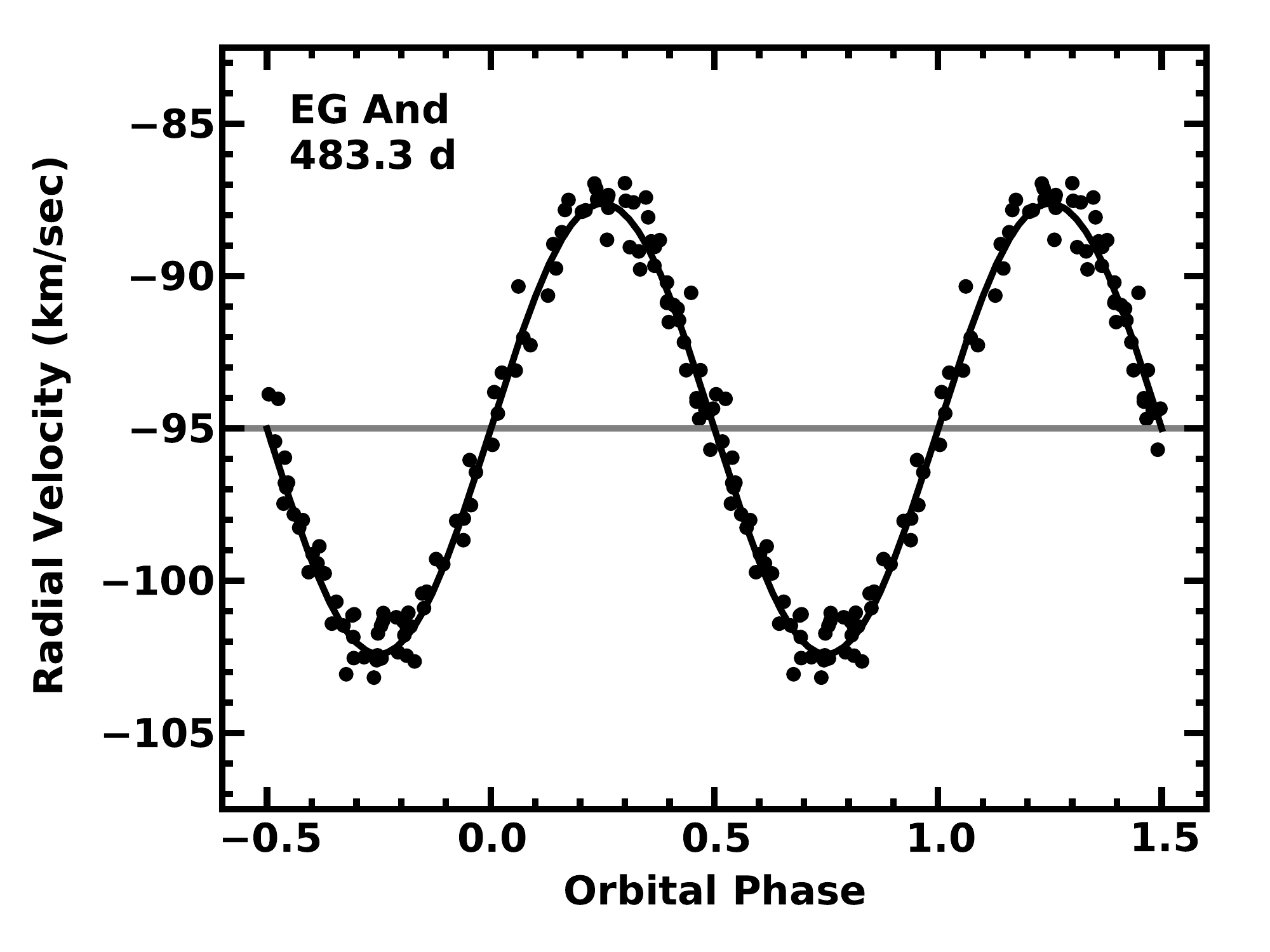}
\vskip 3ex
\caption{
Orbital motion in EG And.  Filled circles: SAO radial velocity data.
Solid black line: best-fitting solution for circular orbit.
Solid horizontal grey line: derived systemic velocity.
\label{fig: orbit}
}
\end{figure}

We analyze the radial velocity data in Table~\ref{tab: rvs} using the
\citet{monet1979} Fourier transform algorithm as implemented by \citet{kg1986}.
To estimate errors in the orbital parameters, we perform a Monte Carlo simulation.
Our approach replaces each velocity measurement $v_i$ with $v_j = v_i + \sigma_v r_j$, 
where $\sigma_v$ is the error in the original measurement (0.75~\kms) and $r_j$ 
is a gaussian random deviate \citep{press1992}. We then derive a best-fitting orbital
period $P$, semi-amplitude $K$, systemic velocity $\gamma$, projected semimajor
axis $a_1$, and time of spectroscopic conjunction $T_0$. Repeat trials yield
a distribution of $N$ values for each orbital parameter. We adopt the median
value as `best' and set the dispersion and inter-quartile range for each parameter 
as the error. In practice, $N = 10^4$ trials provides a robust estimate of the 
period; the dispersion and inter-quartile range are fairly indistinguishable.
For consistency, we choose the inter-quartile range as the error in each parameter.
Fig.~\ref{fig: orbit} plots the data and the orbital solution; Table~\ref{tab: orbs} 
lists our best estimates and errors for the orbital parameters.

The orbital parameters deduced from the SAO data agree well with previous 
spectroscopic estimates.
For an adopted period $P$ = 482~d, \citet{fekel2000} quote a semi-amplitude 
derived from infrared spectra, $K = 7.7 \pm 0.3$~\kms, which differs little 
from our $K$ = 7.34 $\pm$ 0.07~\kms\ for $P$ = 483.3 $\pm$ 1.6~d. Adding 
previous observations from the literature \citep{oliver1985,munari1988,
munari1993}, \citet{fekel2000} quote a final solution with $P$ = 
482.57 $\pm$ 0.53~d and $K$ = 7.32 $\pm$ 0.27~\kms\ that is nearly 
identical to our solution. 

Because \citet{fekel2000} infer orbits using a technique which differs from ours,
we calculate orbital parameters for published data sets using the Fourier transform
algorithm. Table~\ref{tab: orbs} lists results for the \citet{fekel2000} infrared
data (`Fekel') and all previous observations \citep[`OMMF';][]{oliver1985,munari1988,
munari1993,fekel2000}. Within the errors, these solutions agree with those in 
Table~3 of \citet{fekel2000}. Combining these data with our new velocities, we 
derive a combined solution (`All') which differs little from the solution using SAO data
only (`SAO').

Iterating all of these solutions in configuration and Fourier space \citep{kg1986} 
yields a modest orbital eccentricity $e \approx$ 0.02--0.04 which has a formal 
1$\sigma$--2$\sigma$ significance depending on the data set.  For the SAO and 
combined solutions, the \citet{lucy1971} test rules out an eccentric orbit at 
the $\gtrsim 10\sigma$ level \citep[see also][]{bassett1978,lucy1989}. Thus, the 
circular orbital solutions listed in Table~\ref{tab: orbs} are strongly preferred 
\citep[see also the discussions in][]{wilson1997,fekel2000}.

Aside from the eccentricity, these solutions are consistent with previous measurements.
The orbital periods for the SAO and combined solutions, 482.5--483.3 d, are close to
the 482.2 d period inferred from UV eclipses \citep{vogel1991,vogel1992}.
For the combined (SAO) solutions, the time of spectroscopic conjunction, $\rm T_0$ = 
JD~2445384 (JD 2445381) is identical to the epoch of UV photometric minimum, T = 
JD~2445380 \citep{vogel1991}. Our orbital period and time of spectroscopic conjunction 
also agree well with the period and time of minimum inferred from optical photometry 
\citep{skopal1997,wilson1997,skopal2007,jurdana2010,skopal2012}. 

\subsection{Basic System Parameters}
\label{sec: an-pars}

Although previous analyses constrain basic properties of the two 
stellar components, the new spectroscopic orbit and recent advances 
in the effective temperature scale for red giants allows us to 
improve these results.  From a detailed analysis of the UV eclipses, 
\citet{vogel1992} derive $R_g$ = 74 $\pm$ 10~\rsun\ for the radius of 
the giant. The observed K-band brightness and J--K color 
\citep[K = 2.58; J--K = 1.05; e.g.,][]{kg1983,kenyon1988b,phillips2007} 
and the extinction correction \citep[e.g.,][]{savage1979,murset1991} 
then yield an effective temperature, $T_g = 3730 \pm 130$~K \citep{worthey2011};
a bolometric correction at K, BC(K) = $2.7 \pm 0.1$ \citep{worthey2011};
an absolute K-band brightness, M$_K = -5.45 \pm 0.05$; and a revised 
distance of 400 $\pm$ 20~pc. The small formal error follows from the
adopted $\pm$0.05 error in J--K and the relative insensitivity of the 
surface brightness of the red giant as a function of gravity and 
metallicity when log~$g \approx$ 0.5--1 and [Fe/H] $\approx$ 0 
\citep[see][and references therein]{worthey2011}.

This revised distance is identical to a previous estimate from 
\citet{vogel1992}.  Although the nominal Hipparcos distance is 
roughly 700~pc \citep{kolb2004,vanleeuwen2007}, the 1$\sigma$ 
error in the parallax is only slightly smaller than the parallax.
From an analysis of the UV data, \citet{skopal2005} favors roughly 
600~pc. Given the uncertainty in the UV extinction curve and 
the improvements in bolometric corrections for M-type giants, 
we favor a distance of 400~pc.

\begin{figure} 
\includegraphics[width=6.5in]{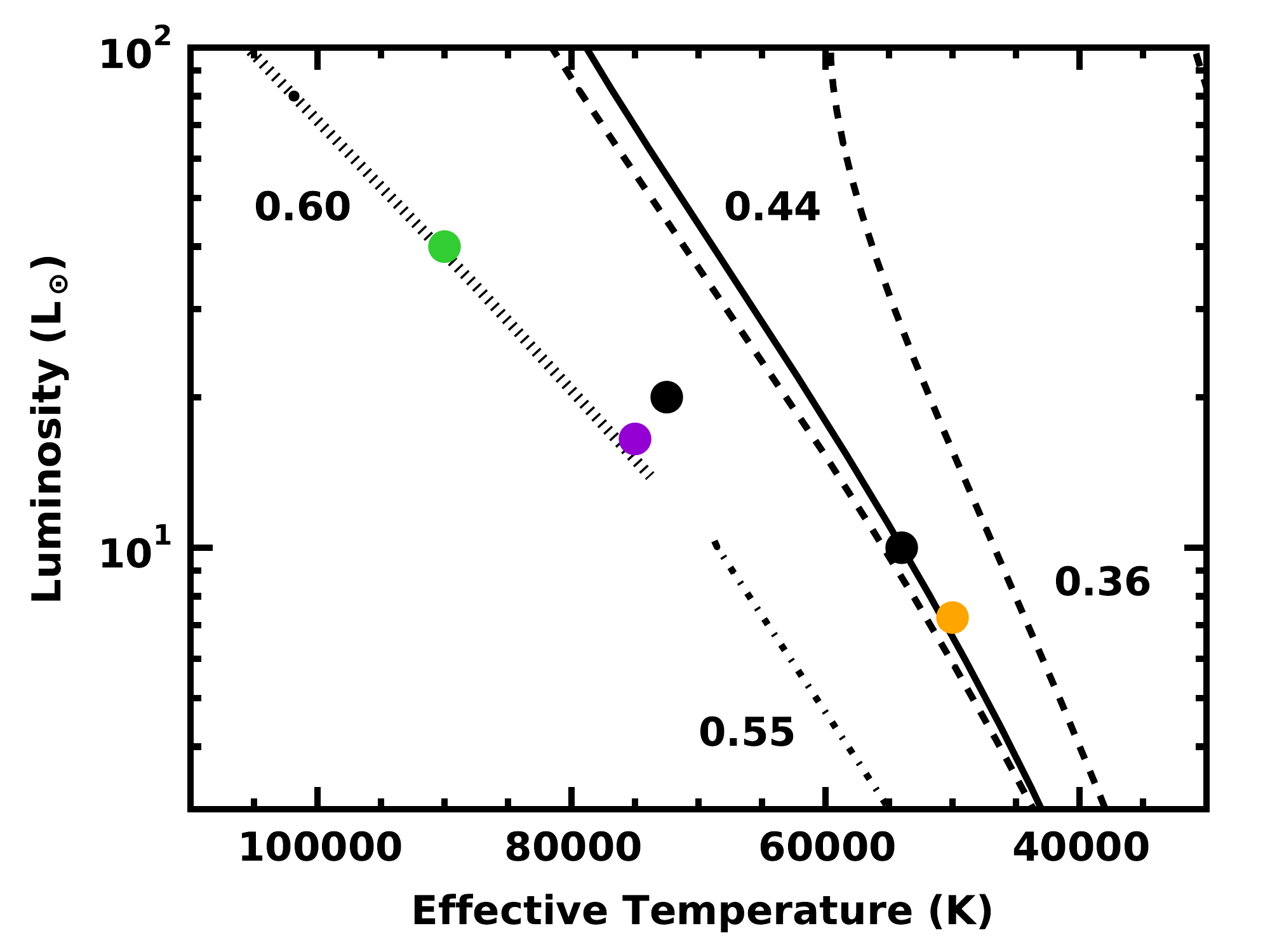}
\vskip 3ex
\caption{
HR diagram for the hot component in EG And.  Filled circles 
indicate published estimates for the effective temperature 
and luminosity from 
\citet[][black]{kenyon1985},
\citet[][purple]{murset1991},
\citet[][orange]{kolb2004}, and
\citet[][green]{skopal2005}.
The scatter provides a measure of the uncertainties 
in the fits to observations of the UV continuum.  
Black curves plot theoretical tracks 
\citep{pac1971,salaris2013,althaus2013} 
for a 0.36~\msun\ He white dwarf (dashed lines),
a 0.44~\msun\ He white dwarf (solid line), a
0.55~\msun\ C--O white dwarf (dot-dashed line),
and a 0.60~\msun\ C--O white dwarf (dotted line).
Multiple lines for the 0.36~\msun\ white dwarf illustrate
the evolution during hydrogen shell flashes. 
\label{fig: hrd}
}
\end{figure}

To estimate the stellar masses, we begin with the position of the hot
star in the HR diagram (Fig.~\ref{fig: hrd}). Previous analyses 
suggest hot component effective temperatures
$T_h \approx$ 50,000--90,000~K \citep{kenyon1985,murset1991,
vogel1992,kolb2004,skopal2005}.  For a 400~pc distance, the 
luminosity is then $L_h \approx$ 5--40~\lsun. The cooler, lower 
luminosity estimates place the hot component close to white dwarf 
cooling curves for 0.36--0.44~\msun\ He white dwarfs \citep{althaus2013}.  
A more luminous hot component falls on cooling curves for 
0.55--0.60~\msun\ C--O white dwarfs \citep{pac1971,salaris2013}. As 
indicated by the multiple tracks for the 0.36~\msun\ white dwarf in 
Fig.~\ref{fig: hrd}, recurrent hydrogen shell flashes can shift the 
cooling curves. Given the uncertainties in the observational estimates 
and the tracks, we adopt a plausible mass range of 0.35--0.55~\msun\ for 
the secondary star.

To constrain properties of the cool primary star, we rely on the derived
mass function and Roche geometry. The mass function relates the component
masses to the orbital semi-amplitude and the orbital period,
$f(M_g, M_h) = (a_1 ~ {\rm sin} ~ i)^3 / P^2$, where 
$a_1 ~ {\rm sin}~i = K P / 2 \pi$.  In terms of the component masses,
\begin{equation}
f(M_g , M_h ) \, = \, \frac{( M_h \, {\rm sin} \, i )^3}{(M_g \, + \, M_h)^2} ~ .
\end{equation} 
When sin~$i \approx$ 1, $M_h \approx (1 + q)^2 f(M_g, M_h)$ where 
$q = M_g / M_h$ is the mass ratio. Once $q$ is known, the size of the 
inner Lagrangian surface sets the maximum radius of the red giant 
\citep{eggle1983}:
\begin{equation}
R_g / a \, \le \frac{0.49 q^{2/3}}{0.6 q^{2/3} \, + \, {\rm ln} \, (1 \, + \, q^{1/3})} \, ,
\end{equation}
where $a$ is the semimajor axis, $a = (1 + q) a_1$.

For $M_h \approx$ 0.35--0.55~\msun, the red giant has a modest range of
allowed masses \citep[see also Table 2 of][]{vogel1992}. The derived 
mass function for the combined set of radial velocities, 
$f(M_g, M_h)$ = 0.019 $\pm$ 0.001, implies $q$ = 3.2--4.4 and 
$M_g \approx$ 1.1--2.4~\msun. As a comparison, radial velocities of the
He II~$\lambda$1640 emission line suggest $q \approx$ 3.5 \citep{crowley2006}.
The maximum radius of the red giant ranges from 
$R_{g, max} \approx$ 185~\rsun\ for $M_h$ = 0.35~\msun\ to 
$R_{g, max} \approx$ 240~\rsun\ for $M_h$ = 0.55~\msun. With an observed radius
of roughly 75\rsun, the giant fills from 40\% ($M_h$ = 0.35~\msun) to
30\% ($M_h$ = 0.55~\msun) of the inner Lagrangian surface.  Our estimate 
for the ratio of the orbital separation to the radius of the red giant, 
$a / R_g \approx$ 3.9--5, compares well with the 
$a / R_g = 4.4 \pm 0.6$ of \citet{vogel1992}.  Given the well-developed 
red giant wind in EG And \citep[e.g.,][]{vogel1992, mika2002}, the lack 
of a lobe-filling giant generally agrees with expectations.

\subsection{Orbital Modulation of H~I emission lines}
\label{sec: an-ew}

\begin{figure} 
\includegraphics[width=6.5in]{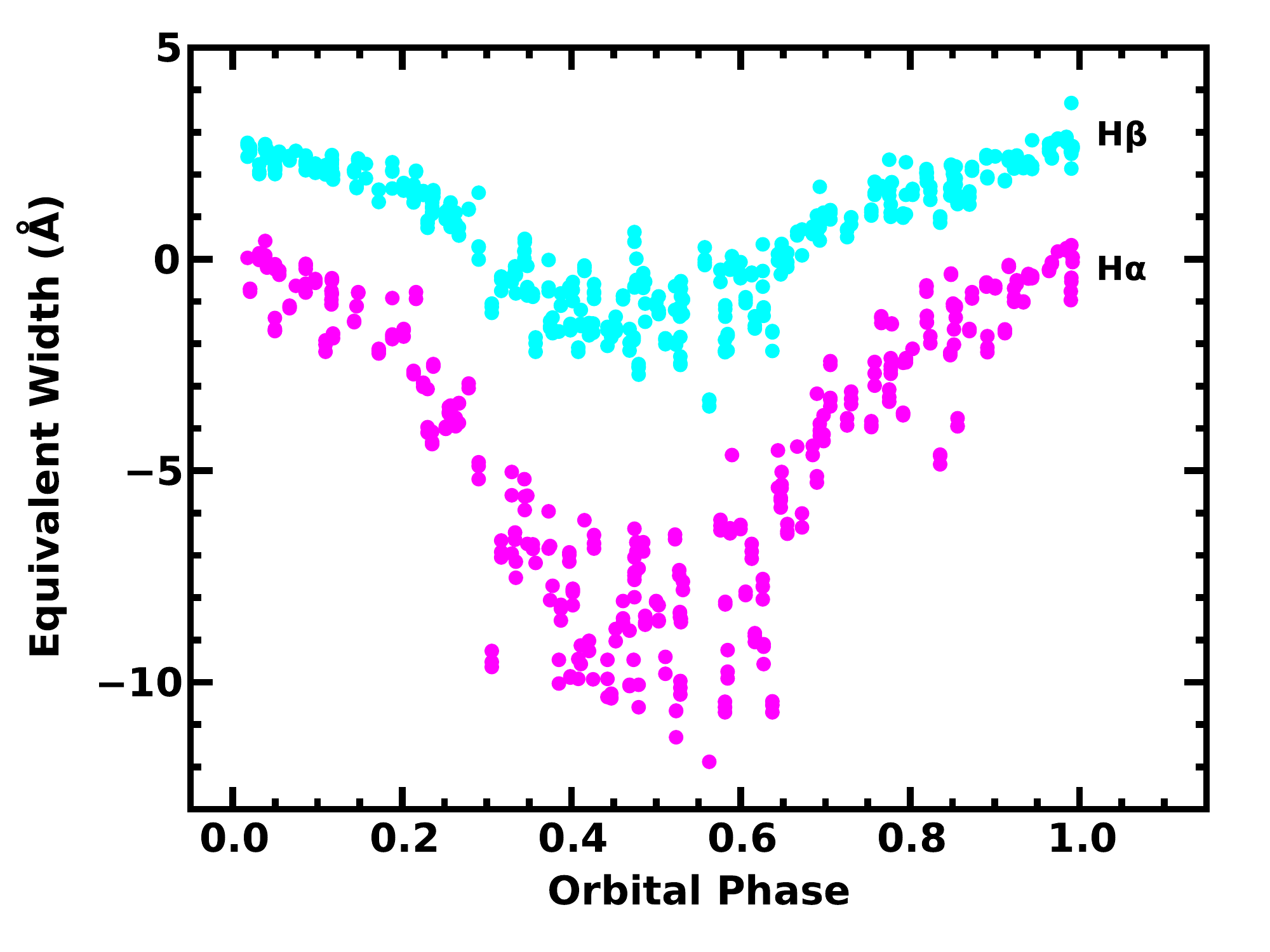}
\vskip 3ex
\caption{
Variation in the measured equivalent widths (EWs) of
H$\alpha$ (magenta points) and H$\beta$ (cyan points)
as a function of orbital phase.  The EWs reach minima 
at $\phi$ = 0.5, when the illuminated hemisphere and
wind of the red giant faces the observer.
\label{fig: lines}
}
\end{figure}

To consider the physical structure of the system in more detail, we now examine the
FAST observations. Fig.~\ref{fig: lines} shows the variation of H$\alpha$ and H$\beta$
as a function of orbital phase. Both lines clearly change in phase with the orbit.
At spectroscopic conjunctions ($\phi$ = 0), the red giant primary lies in front of
the white dwarf secondary.  H$\alpha$ is then a weak emission feature superposed on 
a weak absorption line. The overall EW is zero. As the binary orbits, the emission
component gradually strengthens, reaching peak intensity -- EW $\approx -10\ang$ -- 
when the red giant lies behind the white dwarf. Following this peak, the emission 
component gradually weakens until the EW is once again close to zero at $\phi$ = 1.

The H$\beta$ line behaves in a similar fashion. At $\phi$ = 0, the emission component 
is invisible. The overall EW of 2--3~\ang\ is comparable to the typical EW 
for a normal red giant \citep{oconn1973,montes1997,james2013}. As the white dwarf 
moves in front of the red giant, the emission component increases to a maximum 
intensity with EW $\approx -2~\ang$ at $\phi \approx$ 0.5 and then gradually 
disappears from $\phi \approx$ 0.5 to $\phi \approx$ 1.

Although H$\gamma$ varies in phase with H$\alpha$ and H$\beta$, other emission lines
are too weak to establish coherent time variations. Tests with the strongest He~I, 
[O~II], and [O~III] lines reveal random variations with orbital phase. 
Additional searches failed to find evidence for consistent, detectable emission 
from higher ionization features (e.g., He~II $\lambda$4686 or [Fe~VII] $\lambda$6087).

\begin{figure} 
\includegraphics[width=6.5in]{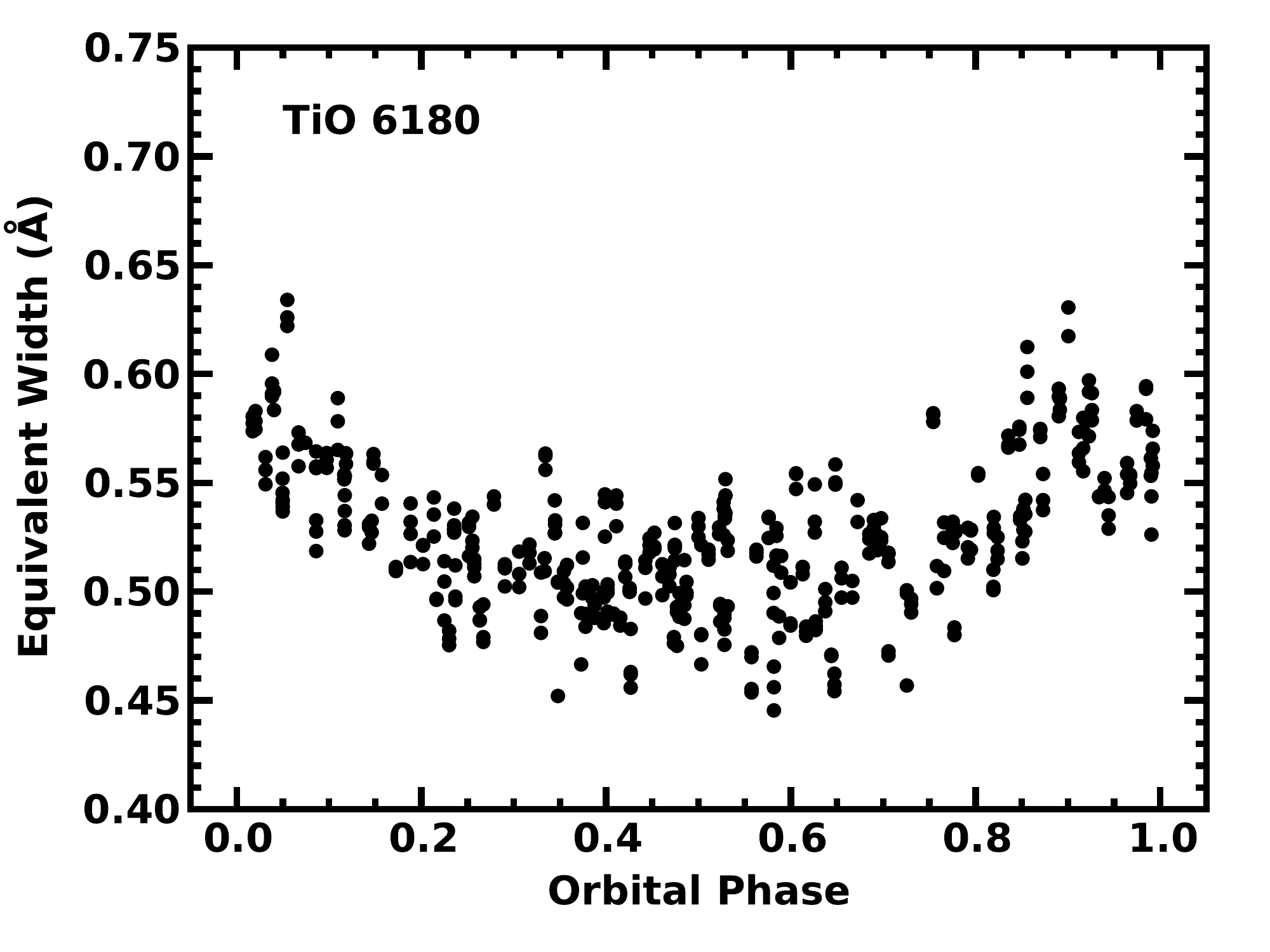}
\vskip 3ex
\caption{
Variation in the measured equivalent width of the
TiO $\lambda$6180 absorption band as a function of 
orbital phase. The absorption is weakest when the
illuminated hemisphere of the red giant faces the
observer.
\label{fig: tio}
}
\end{figure}

Despite this failure, several absorption features clearly change in phase with the 
orbit.  Fig.~\ref{fig: tio} illustrates the variation of the TiO $\lambda$6180
band. Close to $\phi$ = 0, the absorption line is much stronger (EW $\approx$ 
0.55--0.60) than at $\phi \approx$ 0.5 (EW $\approx$ 0.47--0.52). In between
these phases, the absorption gradually strengthens and weakens as the binary
orbits. Other TiO bands and the prominent Ca~I $\lambda$4227 feature similarly
vary in phase with the orbit.

\subsection{Structure of the Emission Line Region}
\label{sec: an-wind}

The time variation of the absorption and emission lines in EG~And is qualitatively 
consistent with the reflection effect. In this interpretation, UV radiation from 
the hot white dwarf secondary illuminates the facing hemisphere of the red giant 
primary. The red giant must then emit a larger luminosity with the same surface 
area, producing a characteristic sinusoidal variation in the optical light curve 
\citep[e.g.,][]{bely1968,kenyon1982,kenyon1984,form1990,skopal2008}. 
The large effective temperature of the more luminous hemisphere weakens various
absorption features, including the TiO bands. Thus, these absorption features 
vary in phase with the optical light curve.

Theoretical models \citep[][1998]{proga1996} demonstrate that high energy photons 
from the hot secondary in a typical symbiotic {\it can} ionize the photosphere and 
the wind from the red giant. Emission line strengths then depend on the UV luminosity 
of the secondary and the physical conditions within the red giant wind. For typical 
parameters, emission line intensities vary in phase with the absorption features 
and optical light curve, reaching maxima at maximum optical brightness and minimum 
absorption line strength.

In their analysis of FUSE and HST STIS data for EG And, \citet{crowley2008} conclude 
that UV emission lines form close to the hot component. Variability in the emission 
line profiles suggests that the hot component accretes material from the red giant 
wind. These data show little or no evidence for a wind from the hot component.  
Coupled with the sub-Eddington luminosity of the hot component,
$L_h / L_{edd} \lesssim 10^{-3}$, the UV and EUV data preclude models where colliding 
winds produce an emission line region between the two stars 
\citep[e.g.,][]{tomov1995,walder1995,calabro2014}. 

Adopting parameters for the hot component derived from fits to UV spectra 
\citep{murset1991} and a density law for the red giant wind \citep{vogel1991}, 
\citet{crowley2008} derive a predicted structure for the ionized nebula 
surrounding the hot component. Aside from a tiny He$^{+2}$ zone, this model 
yields a radius of 85--95~\rsun\ for the size of the H~II region. The H~II 
region is then slightly larger than the 65--85~\rsun\ radius of the M giant.
For $E_{B-V}$ = 0.05 \citep{murset1991}, the predicted\footnote{A simple 
photoionization calculation as in \citet[][1989]{oster1974} and a more 
detailed calculation with Cloudy \citep{ferland2013} yield nearly identical 
results for the expected H$\alpha$ flux.}
H$\alpha$ flux of $1.0 - 1.3 \times 10^{-10}$ erg~cm$^{-2}$~s$^{-1}$ 
is within 5\%--10\% of the maximum flux observed at $\phi$ = 0.5.

The orbital variations of H$\alpha$ and H$\beta$ on FAST spectra are consistent
with this picture.  In the \citet{crowley2008} ionization model, we expect to 
see roughly 20\% of the H$\alpha$ emission at mid-eclipse when the giant completely 
occults the hot component.  The FAST observations suggest $| {\rm EW } | \approx$ 
1--2~\ang\ for H$\alpha$ emission at $\phi \approx$ 0 and 9--11~\ang\ at 
$\phi$ = 0.5. Thus, roughly 10\%--20\% of the H$\alpha$ emission is visible at 
mid-eclipse, agreeing with expectations. For H$\beta$, the orbital modulation 
suggests that 5\%--10\% of the H~II region is visible, consistent with standard
photoionization models where H$\alpha$ forms in a slightly larger region than
H$\beta$ \citep[e.g.,][]{oster1989,ferland2013}.

Orbit-to-orbit fluctuations in H$\alpha$ and H$\beta$ emission at 
$\phi$ = 0.5 are also compatible with the FUSE and STIS observations.  
As discussed in \citet{crowley2008}, variations in the fluxes and 
profiles of strong emission lines suggest changes in the accretion 
rate onto the hot component. The FAST observations suggest 20\% to 
40\% fluctuations in H$\alpha$, compared to 20\% to 50\% variations 
in the far-UV continuum and emission lines. 

Although \citet{crowley2008} conclude that the hot component has little 
impact on the absorption spectrum of the red giant, the variations 
in the TiO absorption index are qualitatively consistent with the 
ionized wind model.  Aside from H$\alpha$, the ionized wind emits 
Balmer and Paschen continuum radiation \citep[e.g.,][]{oster1989}. 
The flux from this component should track the flux from H$\alpha$.  
Thus, weaker (stronger) H$\alpha$ emission implies a weaker (stronger) 
Paschen continuum and a stronger (weaker) TiO absorption band, as 
observed.  However, photoionization models predict that the nebula 
contributes less than 0.01\% of the continuum flux at 6000--7000~\ang, 
much smaller than the amplitude of the variation in the red TiO bands. 

Changes in the photospheric temperature of the red giant seem a more
likely source of variability in the absorption features. The observed
10\% fluctuations in TiO absorption correspond to 0.5 subclass 
($\lesssim$ 50~K) changes in the spectral type (effective temperature)
of the giant \citep{kenyon1987a}. Emission from the hot component and
the ionized wind may be sufficient to raise the temperature of the
red giant in the hemisphere which faces the hot component 
\citep[e.g.,][]{proga1998}.

\subsection{Disk Accretion}
\label{sec: an-acc}

With little evidence for colliding winds in the system,
it is worth considering whether the hot white dwarf accretes
material from the red giant wind. Based on the shape and time
variation of O~VI $\lambda\lambda$1032, 1036 line profiles,
\citet{crowley2008} conclude that some accretion is likely.  
In their analysis of EG And's X-ray emission, \citet{nunez2015}
infer an X-ray luminosity, $L_X \approx 5 \times 10^{-4} \lsun$,
for optically thin material close to the white dwarf. For
a white dwarf with mass $M_{wd} \approx$ 0.4--0.5~\msun, 
radius $R_{wd} \approx 10^9$~cm \citep{prove1998,holberg2012}, 
and X-ray luminosity 
$L_X = G M_{wd} \dot{M}_{wd} / 2 R_{wd}$, an accretion rate
$\dot{M}_{wd} \approx 10^{-12}$~\msunyr\ can power the observed
X-ray flux. This rate is 0.01\% of the mass loss rate of the 
red giant, $\dot{M}_g \approx 10^{-8} \msunyr$ 
\citep{vogel1991,skt1993,mika2002}.

To generate this accretion rate, geometric capture is insufficent.  
Adopting $\dot{M}_{wd} \approx (R_{wd}^2 / 4 a^2) \dot{M}_g$, a 
typical white dwarf radius, and the measured $a$ yields 
$\dot{M}_{wd} \approx 10^{-17} \msunyr$. Thus, substantial gravitational
focusing is necessary \citep[see also][]{shore2010,shagatova2016}.

In the Bondi-Hoyle formalism, disk accretion occurs when the
specific angular momentum of accreted material exceeds the 
specific angular momentum of material orbiting the equator
of the white dwarf \citep[e.g.,][]{soker2000,perets2013}. The
accretion rate then depends on the Bondi-Hoyle accretion radius,
\begin{equation}
R_a = {2 G M_h \over v_r^2 + c_s^2 } ,
\end{equation}
where $v_r$ is the velocity of material near the white dwarf and
$c_s$ is the sound speed.  Adopting a reasonable range for $v_r$ 
= 30--80~\kms\ and $c_s$ = 10--20~\kms, the accretion rate is then 
$\dot{M}_{wd} \approx (R_a / 2 a)^2 \dot{M}_g$ $\approx$ 
0.003--0.05~$\dot{M}_g$ $\approx$ $3 - 50 \times 10^{-11}$~\msunyr.  
Even though the red giant fills only 30\%--40\% of its Roche lobe, 
the Bondi-Hoyle rate is still 30--500 times larger than the rate 
required to power the X-ray flux.  Thus, accretion can easily power 
the observed X-ray flux.  With a total accretion luminosity of only 
0.03--0.5~\lsun, however, the energy from Bondi-Hoyle accretion is 
insufficient to power the observed UV luminosity of the hot white 
dwarf.  More rigorous numerical models \citep[e.g.,][]{dvb2009} 
are required to infer the properties of the accreted material in 
more detail.

\section{SUMMARY}
\label{sec: disc}

Our analysis of optical spectroscopic observations places better 
constraints on the physical structure of the EG~And binary.  Roughly 
100 new radial velocities yield improved parameters for a circular
orbit and eliminate eccentric orbits at the 10$\sigma$ level.
Measurements of orbital variations in H$\alpha$ yield a strong test 
of illumination models, where high energy photons from the hot
component ionize the wind from the red giant. When the H$\alpha$ 
emission line reaches maximum intensity, the observed flux is 
within 10\% of predictions. The small emission flux during the 
eclipse of the hot component is also consistent with predictions.

Together with an absolute measure of the red giant radius 
\citep{vogel1991}, updated data for effective temperatures and 
bolometric corrections \citep{worthey2011} yield a distance, 
400~pc, which differs little from previous estimates.  For this 
distance, the luminosity and effective temperature of the hot 
component are consistent with theoretical models for 
0.35--0.55~\msun\ He/C--O white dwarfs with H envelopes.  The 
revised spectroscopic mass function then implies a red giant 
mass $M_g \approx$ 1.1--2.5~\msun.
The large mass range for the hot component results mainly from 
factor of roughly 2 uncertainties in the effective temperature 
and a factor of four uncertainty in the luminosity. Now that 
H$\alpha$ observations confirm aspects of the illumination model, 
detailed comparisons between predictions and observations of EUV 
and UV emission lines might yield better constraints on the 
effective temperature.

Understanding the orbital modulation of TiO and other strong
absorption features requires high resolution spectra covering 
a large wavelength range (e.g., 3500--8000~\ang). Data at 
3500--3800~\ang\ might reveal a Balmer emission jump, which 
could be used to constrain the continuum from the photoionized 
wind. At longer wavelengths, profile variations in Ca~I and
other atomic lines along with equivalent width variations in
TiO and other molecular bands could establish any variation
of effective temperature or gravity around the orbit.

Another uncertainty in the system geometry is the possible 
ellipsoidal variation of the primary star \citep{wilson1997}. 
The existence of this variability is somewhat controversial 
\citep{walder1995,tomov1995,skopal2007,skopal2008,jurdana2010,
skopal2012,calabro2014}.  Moreover, light curve fits yield a 
red giant which fills 85\%--97\% of its tidal surface; the 
derived red giant radii, $R_g \approx$ 100--230~\rsun\ for 
$i \approx$ 45\deg--70\deg, are then much larger than the 
$R_g = 74 \pm 10$~\rsun\ inferred from the UV eclipses. A large
red giant also seems inconsistent with the existence of a 
fairly smooth red giant wind and the apparent lack of a large,
luminous accretion disk \citep{vogel1991,kolb2004}. Ellipsoidal 
light curve analyses on larger sets of optical photometry assuming
a more edge-on geometry (80\deg--90\deg) might allow a more 
consistent solution or completely eliminate this possibility.

In the next few years, Gaia observations will resolve the uncertainties 
in distance estimates \citep{linde2010}. For stars as optically bright 
as EG And with distances smaller than 1~kpc, \citet{linde2012} predict 
parallaxes with 1\% or better accuracy. Once Gaia releases improved
distances for EG And and other nearby symbiotics, the luminosities of
the hot components will only be sensitive to uncertain effective 
temperatures. If improved modeling techniques yield smaller uncertainties
in the effective temperatures, these distances would enable direct
comparisons of the properties of the hot components with theoretical
tracks of white dwarfs in the HR diagram.

\acknowledgements
We thank D. Proga for helpful discussions and comments that improved 
the manuscript. Comments from an anonymous reviewer also honed our
discussion.
We acknowledge a generous allotment of telescope time on the SAO 
1.5-m telescopes and the NOAO 0.9-m telescopes.  This paper uses data 
products produced by the OIR Telescope Data Center, supported by the 
Smithsonian Astrophysical Observatory.  The photoionization calculations 
used version 13.03 of Cloudy, last described in \citet{ferland2013}.

{\it Facility:} \facility{KPNO:0.9m, FLWO:1.5m, ORO:1.5m}
%\bibliography{ss}
\bibliography{ms.bbl}

\begin{thebibliography}{87}
\expandafter\ifx\csname natexlab\endcsname\relax\def\natexlab#1{#1}\fi

\bibitem[{{Allen}(1984)}]{allen1984}
{Allen}, D.~A. 1984, Proceedings of the Astronomical Society of Australia, 5,
  369

\bibitem[{{Althaus} {et~al.}(2013){Althaus}, {Miller Bertolami}, \&
  {C{\'o}rsico}}]{althaus2013}
{Althaus}, L.~G., {Miller Bertolami}, M.~M., \& {C{\'o}rsico}, A.~H. 2013,
  \aap, 557, A19

\bibitem[{{Barnes} \& {Hayes}(1984)}]{barnes1984}
{Barnes}, J.~V., \& {Hayes}, D.~S. 1984, {IRS Standard Star Manual}
  (Association of Universities for Research in Astronomy, Tucson, AZ USA)

\bibitem[{{Bassett}(1978)}]{bassett1978}
{Bassett}, E.~E. 1978, The Observatory, 98, 122

\bibitem[{{Belczy{\'n}ski} {et~al.}(2000){Belczy{\'n}ski}, {Miko{\l}ajewska},
  {Munari}, {Ivison}, \& {Friedjung}}]{belczy2000}
{Belczy{\'n}ski}, K., {Miko{\l}ajewska}, J., {Munari}, U., {Ivison}, R.~J., \&
  {Friedjung}, M. 2000, \aaps, 146, 407

\bibitem[{{Belyakina}(1968)}]{bely1968}
{Belyakina}, T.~S. 1968, \azh, 45, 139

\bibitem[{{Calabr{\`o}}(2014)}]{calabro2014}
{Calabr{\`o}}, E. 2014, Journal of Astrophysics and Astronomy, 35, 69

\bibitem[{{Chakrabarty} \& {Roche}(1997)}]{chak1997}
{Chakrabarty}, D., \& {Roche}, P. 1997, \apj, 489, 254

\bibitem[{{Corradi} {et~al.}(2008){Corradi}, {Rodr{\'{\i}}guez-Flores},
  {Mampaso}, {Greimel}, {Viironen}, {Drew}, {Lennon}, {Mikolajewska}, {Sabin},
  \& {Sokoloski}}]{corradi2008}
{Corradi}, R.~L.~M., {et~al.} 2008, \aap, 480, 409

\bibitem[{{Crowley}(2006)}]{crowley2006}
{Crowley}, C. 2006, PhD thesis, School of Physics, Trinity College Dublin,
  Dublin 2, Ireland (https://www.tcd.ie/Physics/Astrophysics/crowley.php)

\bibitem[{{Crowley} {et~al.}(2008){Crowley}, {Espey}, \&
  {McCandliss}}]{crowley2008}
{Crowley}, C., {Espey}, B.~R., \& {McCandliss}, S.~R. 2008, \apj, 675, 711

\bibitem[{{de Val-Borro} {et~al.}(2009){de Val-Borro}, {Karovska}, \&
  {Sasselov}}]{dvb2009}
{de Val-Borro}, M., {Karovska}, M., \& {Sasselov}, D. 2009, \apj, 700, 1148

\bibitem[{{Eggleton}(1983)}]{eggle1983}
{Eggleton}, P.~P. 1983, \apj, 268, 368

\bibitem[{{Fabricant} {et~al.}(1998){Fabricant}, {Cheimets}, {Caldwell}, \&
  {Geary}}]{fab1998}
{Fabricant}, D., {Cheimets}, P., {Caldwell}, N., \& {Geary}, J. 1998, \pasp,
  110, 79

\bibitem[{{Fekel} {et~al.}(2000){Fekel}, {Joyce}, {Hinkle}, \&
  {Skrutskie}}]{fekel2000}
{Fekel}, F.~C., {Joyce}, R.~R., {Hinkle}, K.~H., \& {Skrutskie}, M.~F. 2000,
  \aj, 119, 1375

\bibitem[{{Ferland} {et~al.}(2013){Ferland}, {Porter}, {van Hoof}, {Williams},
  {Abel}, {Lykins}, {Shaw}, {Henney}, \& {Stancil}}]{ferland2013}
{Ferland}, G.~J., {et~al.} 2013, \rmxaa, 49, 137

\bibitem[{{Formiggini} \& {Leibowitz}(1990)}]{form1990}
{Formiggini}, L., \& {Leibowitz}, E.~M. 1990, \aap, 227, 121

\bibitem[{{Garcia}(1986)}]{garcia1986}
{Garcia}, M.~R. 1986, \aj, 91, 1400

\bibitem[{{Garcia} \& {Kenyon}(1988)}]{garcia1988}
{Garcia}, M.~R., \& {Kenyon}, S.~J. 1988, in IAU Colloq.~103: The Symbiotic
  Phenomenon, ed. J.~{Mikolajewska}, M.~{Friedjung}, S.~J. {Kenyon}, \&
  R.~{Viotti}, 27

\bibitem[{{Hartmann} {et~al.}(1986){Hartmann}, {Hewett}, {Stahler}, \&
  {Mathieu}}]{hartmann1986}
{Hartmann}, L., {Hewett}, R., {Stahler}, S., \& {Mathieu}, R.~D. 1986, \apj,
  309, 275

\bibitem[{{Hayes} \& {Latham}(1975)}]{hayes1975}
{Hayes}, D.~S., \& {Latham}, D.~W. 1975, \apj, 197, 593

\bibitem[{{Hinkle} {et~al.}(2006){Hinkle}, {Fekel}, {Joyce}, {Wood}, {Smith},
  \& {Lebzelter}}]{hinkle2006}
{Hinkle}, K.~H., {Fekel}, F.~C., {Joyce}, R.~R., {Wood}, P.~R., {Smith}, V.~V.,
  \& {Lebzelter}, T. 2006, \apj, 641, 479

\bibitem[{{Holberg} {et~al.}(2012){Holberg}, {Oswalt}, \&
  {Barstow}}]{holberg2012}
{Holberg}, J.~B., {Oswalt}, T.~D., \& {Barstow}, M.~A. 2012, \aj, 143, 68

\bibitem[{{James}(2013)}]{james2013}
{James}, D.~J. 2013, \pasp, 125, 1087

\bibitem[{{Jurdana-{\v S}epi{\'c}} \& {Munari}(2010)}]{jurdana2010}
{Jurdana-{\v S}epi{\'c}}, R., \& {Munari}, U. 2010, \pasp, 122, 35

\bibitem[{{Kaler} \& {Hickey}(1983)}]{kaler1983}
{Kaler}, J.~B., \& {Hickey}, J.~P. 1983, \pasp, 95, 759

\bibitem[{{Kenyon}(1982)}]{kenyon1982}
{Kenyon}, S.~J. 1982, \pasp, 94, 165

\bibitem[{{Kenyon}(1985)}]{kenyon1985}
{Kenyon}, S.~J. 1985, in Astrophysics and Space Science Library, Vol. 113,
  Cataclysmic Variables and Low-Mass X-ray Binaries, ed. D.~Q. {Lamb} \&
  J.~{Patterson}, 417--423

\bibitem[{{Kenyon}(1986)}]{kenyon1986}
---. 1986, {The symbiotic stars} (Cambridge University Press, Cambridge, UK)

\bibitem[{{Kenyon}(1988)}]{kenyon1988b}
---. 1988, \aj, 96, 337

\bibitem[{{Kenyon} \& {Bateson}(1984)}]{kenyon1984}
{Kenyon}, S.~J., \& {Bateson}, F.~M. 1984, \pasp, 96, 321

\bibitem[{{Kenyon} \& {Fernandez-Castro}(1987)}]{kenyon1987a}
{Kenyon}, S.~J., \& {Fernandez-Castro}, T. 1987, \aj, 93, 938

\bibitem[{{Kenyon} {et~al.}(1988){Kenyon}, {Fernandez-Castro}, \&
  {Stencel}}]{kenyon1988a}
{Kenyon}, S.~J., {Fernandez-Castro}, T., \& {Stencel}, R.~E. 1988, \aj, 95,
  1817

\bibitem[{{Kenyon} \& {Gallagher}(1983)}]{kg1983}
{Kenyon}, S.~J., \& {Gallagher}, J.~S. 1983, \aj, 88, 666

\bibitem[{{Kenyon} \& {Garcia}(1986)}]{kg1986}
{Kenyon}, S.~J., \& {Garcia}, M.~R. 1986, \aj, 91, 125

\bibitem[{{Kenyon} {et~al.}(1991){Kenyon}, {Oliversen}, {Mikolajewska},
  {Mikolajewski}, {Stencel}, {Garcia}, \& {Anderson}}]{kenyon1991}
{Kenyon}, S.~J., {Oliversen}, N.~A., {Mikolajewska}, J., {Mikolajewski}, M.,
  {Stencel}, R.~E., {Garcia}, M.~R., \& {Anderson}, C.~M. 1991, \aj, 101, 637

\bibitem[{{Kenyon} \& {Webbink}(1984)}]{kw1984}
{Kenyon}, S.~J., \& {Webbink}, R.~F. 1984, \apj, 279, 252

\bibitem[{{Kolb} {et~al.}(2004){Kolb}, {Miller}, {Sion}, \&
  {Miko{\l}ajewska}}]{kolb2004}
{Kolb}, K., {Miller}, J., {Sion}, E.~M., \& {Miko{\l}ajewska}, J. 2004, \aj,
  128, 1790

\bibitem[{{Latham}(1985)}]{latham1985}
{Latham}, D.~W. 1985, in Stellar Radial Velocities, ed. A.~G.~D. {Philip} \&
  D.~W. {Latham}, 21--34

\bibitem[{{Lindegren}(2010)}]{linde2010}
{Lindegren}, L. 2010, in IAU Symposium, Vol. 261, IAU Symposium, ed. S.~A.
  {Klioner}, P.~K. {Seidelmann}, \& M.~H. {Soffel}, 296--305

\bibitem[{{Lindegren} {et~al.}(2012){Lindegren}, {Lammers}, {Hobbs},
  {O'Mullane}, {Bastian}, \& {Hern{\'a}ndez}}]{linde2012}
{Lindegren}, L., {Lammers}, U., {Hobbs}, D., {O'Mullane}, W., {Bastian}, U., \&
  {Hern{\'a}ndez}, J. 2012, \aap, 538, A78

\bibitem[{{Lucy}(1989)}]{lucy1989}
{Lucy}, L.~B. 1989, The Observatory, 109, 100

\bibitem[{{Lucy} \& {Sweeney}(1971)}]{lucy1971}
{Lucy}, L.~B., \& {Sweeney}, M.~A. 1971, \aj, 76, 544

\bibitem[{{Massey} {et~al.}(1988){Massey}, {Strobel}, {Barnes}, \&
  {Anderson}}]{massey1988}
{Massey}, P., {Strobel}, K., {Barnes}, J.~V., \& {Anderson}, E. 1988, \apj,
  328, 315

\bibitem[{{Mazeh} {et~al.}(1996){Mazeh}, {Latham}, \& {Stefanik}}]{mazeh1996}
{Mazeh}, T., {Latham}, D.~W., \& {Stefanik}, R.~P. 1996, \apj, 466, 415

\bibitem[{{Miko{\l}ajewska} {et~al.}(2002){Miko{\l}ajewska}, {Ivison}, \&
  {Omont}}]{mika2002}
{Miko{\l}ajewska}, J., {Ivison}, R.~J., \& {Omont}, A. 2002, Advances in Space
  Research, 30, 2045

\bibitem[{{Monet}(1979)}]{monet1979}
{Monet}, D.~G. 1979, \apj, 234, 275

\bibitem[{{Montes} {et~al.}(1997){Montes}, {Martin}, {Fernandez-Figueroa},
  {Cornide}, \& {de Castro}}]{montes1997}
{Montes}, D., {Martin}, E.~L., {Fernandez-Figueroa}, M.~J., {Cornide}, M., \&
  {de Castro}, E. 1997, \aaps, 123

\bibitem[{{Muerset} {et~al.}(1991){Muerset}, {Nussbaumer}, {Schmid}, \&
  {Vogel}}]{murset1991}
{Muerset}, U., {Nussbaumer}, H., {Schmid}, H.~M., \& {Vogel}, M. 1991, \aap,
  248, 458

\bibitem[{{Munari}(1993)}]{munari1993}
{Munari}, U. 1993, \aap, 273, 425

\bibitem[{{Munari} {et~al.}(1988){Munari}, {Margoni}, {Iijima}, \&
  {Mammano}}]{munari1988}
{Munari}, U., {Margoni}, R., {Iijima}, T., \& {Mammano}, A. 1988, \aap, 198,
  173

\bibitem[{{Nu{\~n}ez} {et~al.}(2015){Nu{\~n}ez}, {Nelson}, {Mukai},
  {Sokoloski}, \& {Luna}}]{nunez2015}
{Nu{\~n}ez}, N.~E., {Nelson}, T., {Mukai}, K., {Sokoloski}, J.~L., \& {Luna},
  G.~J.~M. 2015, ArXiv e-prints

\bibitem[{{O'Connell}(1973)}]{oconn1973}
{O'Connell}, R.~W. 1973, \aj, 78, 1074

\bibitem[{{Oliversen} {et~al.}(1985){Oliversen}, {Anderson}, {Slovak}, \&
  {Stencel}}]{oliver1985}
{Oliversen}, N.~A., {Anderson}, C.~M., {Slovak}, M.~H., \& {Stencel}, R.~E.
  1985, \apj, 295, 620

\bibitem[{{Osterbrock}(1974)}]{oster1974}
{Osterbrock}, D.~E. 1974, {Astrophysics of gaseous nebulae} (W.~H.~Freeman and
  Co., San Francisco, CA)

\bibitem[{{Osterbrock}(1989)}]{oster1989}
---. 1989, {Astrophysics of gaseous nebulae and active galactic nuclei}
  (University Science Books, Mill Valley, CA)

\bibitem[{{Paczy{\'n}ski}(1971)}]{pac1971}
{Paczy{\'n}ski}, B. 1971, \actaa, 21, 417

\bibitem[{{Perets} \& {Kenyon}(2013)}]{perets2013}
{Perets}, H.~B., \& {Kenyon}, S.~J. 2013, \apj, 764, 169

\bibitem[{{Phillips}(2007)}]{phillips2007}
{Phillips}, J.~P. 2007, \mnras, 376, 1120

\bibitem[{{Press} {et~al.}(1992){Press}, {Teukolsky}, {Vetterling}, \&
  {Flannery}}]{press1992}
{Press}, W.~H., {Teukolsky}, S.~A., {Vetterling}, W.~T., \& {Flannery}, B.~P.
  1992, {Numerical recipes in C. The art of scientific computing} (Cambridge:
  University Press, Cambridge, UK)

\bibitem[{{Proga} {et~al.}(1998){Proga}, {Kenyon}, \& {Raymond}}]{proga1998}
{Proga}, D., {Kenyon}, S.~J., \& {Raymond}, J.~C. 1998, \apj, 501, 339

\bibitem[{{Proga} {et~al.}(1996){Proga}, {Kenyon}, {Raymond}, \&
  {Mikolajewska}}]{proga1996}
{Proga}, D., {Kenyon}, S.~J., {Raymond}, J.~C., \& {Mikolajewska}, J. 1996,
  \apj, 471, 930

\bibitem[{{Provencal} {et~al.}(1998){Provencal}, {Shipman}, {H{\o}g}, \&
  {Thejll}}]{prove1998}
{Provencal}, J.~L., {Shipman}, H.~L., {H{\o}g}, E., \& {Thejll}, P. 1998, \apj,
  494, 759

\bibitem[{{Salaris} {et~al.}(2013){Salaris}, {Althaus}, \&
  {Garc{\'{\i}}a-Berro}}]{salaris2013}
{Salaris}, M., {Althaus}, L.~G., \& {Garc{\'{\i}}a-Berro}, E. 2013, \aap, 555,
  A96

\bibitem[{{Savage} \& {Mathis}(1979)}]{savage1979}
{Savage}, B.~D., \& {Mathis}, J.~S. 1979, \araa, 17, 73

\bibitem[{{Seaquist} {et~al.}(1993){Seaquist}, {Krogulec}, \&
  {Taylor}}]{skt1993}
{Seaquist}, E.~R., {Krogulec}, M., \& {Taylor}, A.~R. 1993, \apj, 410, 260

\bibitem[{{Shagatova} {et~al.}(2016){Shagatova}, {Skopal}, \&
  {Carikov{\'a}}}]{shagatova2016}
{Shagatova}, N., {Skopal}, A., \& {Carikov{\'a}}, Z. 2016, \aap, 588, A83

\bibitem[{{Shore} \& {Wahlgren}(2010)}]{shore2010}
{Shore}, S.~N., \& {Wahlgren}, G.~M. 2010, \aap, 515, A108

\bibitem[{{Skopal}(1997)}]{skopal1997}
{Skopal}, A. 1997, \aap, 318, 53

\bibitem[{{Skopal}(2005)}]{skopal2005}
{Skopal}, A. 2005, in Astronomical Society of the Pacific Conference Series,
  Vol. 330, The Astrophysics of Cataclysmic Variables and Related Objects, ed.
  J.-M. {Hameury} \& J.-P. {Lasota}, 463

\bibitem[{{Skopal}(2008)}]{skopal2008}
---. 2008, Journal of the American Association of Variable Star Observers
  (JAAVSO), 36, 9

\bibitem[{{Skopal} {et~al.}(1991){Skopal}, {Chochol}, {Vittone}, {Blanco}, \&
  {Mammano}}]{skopal1991}
{Skopal}, A., {Chochol}, D., {Vittone}, A.~A., {Blanco}, C., \& {Mammano}, A.
  1991, \aap, 245, 531

\bibitem[{{Skopal} {et~al.}(2012){Skopal}, {Shugarov}, {Va{\v n}ko},
  {Dubovsk{\'y}}, {Peneva}, {Semkov}, \& {Wolf}}]{skopal2012}
{Skopal}, A., {Shugarov}, S., {Va{\v n}ko}, M., {Dubovsk{\'y}}, P., {Peneva},
  S.~P., {Semkov}, E., \& {Wolf}, M. 2012, Astronomische Nachrichten, 333, 242

\bibitem[{{Skopal} {et~al.}(2007){Skopal}, {Va{\v n}ko}, {Pribulla}, {Chochol},
  {Semkov}, {Wolf}, \& {Jones}}]{skopal2007}
{Skopal}, A., {Va{\v n}ko}, M., {Pribulla}, T., {Chochol}, D., {Semkov}, E.,
  {Wolf}, M., \& {Jones}, A. 2007, Astronomische Nachrichten, 328, 909

\bibitem[{{Soker} \& {Rappaport}(2000)}]{soker2000}
{Soker}, N., \& {Rappaport}, S. 2000, \apj, 538, 241

\bibitem[{{Sokoloski} {et~al.}(2006){Sokoloski}, {Kenyon}, {Espey}, {Keyes},
  {McCandliss}, {Kong}, {Aufdenberg}, {Filippenko}, {Li}, {Brocksopp},
  {Kaiser}, {Charles}, {Rupen}, \& {Stone}}]{sokol2006}
{Sokoloski}, J.~L., {et~al.} 2006, \apj, 636, 1002

\bibitem[{{Stencel}(1984)}]{stencel1984}
{Stencel}, R.~E. 1984, \apjl, 281, L75

\bibitem[{{Tomov} \& {Tomova}(1996)}]{tomov1996}
{Tomov}, N., \& {Tomova}, M. 1996, Information Bulletin on Variable Stars,
  4341, 1

\bibitem[{{Tomov}(1995)}]{tomov1995}
{Tomov}, N.~A. 1995, \mnras, 272, 189

\bibitem[{{Tonry} \& {Davis}(1979)}]{tonry1979}
{Tonry}, J., \& {Davis}, M. 1979, \aj, 84, 1511

\bibitem[{{van Leeuwen}(2007)}]{vanleeuwen2007}
{van Leeuwen}, F. 2007, \aap, 474, 653

\bibitem[{{Vogel}(1991)}]{vogel1991}
{Vogel}, M. 1991, \aap, 249, 173

\bibitem[{{Vogel} {et~al.}(1992){Vogel}, {Nussbaumer}, \& {Monier}}]{vogel1992}
{Vogel}, M., {Nussbaumer}, H., \& {Monier}, R. 1992, \aap, 260, 156

\bibitem[{{Walder}(1995)}]{walder1995}
{Walder}, R. 1995, in Annals of the Israel Physical Society, Vol.~11,
  Asymmetrical Planetary Nebulae, ed. A.~{Harpaz} \& N.~{Soker}, 248

\bibitem[{{Wilson}(1950)}]{wilson1950}
{Wilson}, R.~E. 1950, \pasp, 62, 14

\bibitem[{{Wilson} \& {Vaccaro}(1997)}]{wilson1997}
{Wilson}, R.~E., \& {Vaccaro}, T.~R. 1997, \mnras, 291, 54

\bibitem[{{Worthey} \& {Lee}(2011)}]{worthey2011}
{Worthey}, G., \& {Lee}, H.-C. 2011, \apjs, 193, 1

\end{thebibliography}

\clearpage

\begin{deluxetable}{lcrrclcrr}
\tablecolumns{9}
\tablecaption{EG And Radial Velocity Data}
\tabletypesize{\footnotesize}
\tablehead{
{MJD} & {~~~Phase~~~} & 
{~~~~$v_r$~~~~~} & {~~~~~~O--C~~} & 
{~~} &
{MJD} & {~~~Phase~~~} & {~~~$v_r$~~~~} & {~~~~~O--C~~} }
\startdata
45243.7815 & 0.717 & $-$102.51 &  0.45 & & 48310.4657 & 0.062 & $-$90.34  & $-$1.75 \\
45308.5671 & 0.851 & $-$100.90 &  0.11 & & 48434.8081 & 0.319 & $-$87.58  & $-$0.66 \\
45507.8122 & 0.263 &  $-$87.76 &  0.18 & & 48457.8627 & 0.367 & $-$89.04  & $-$0.41 \\
45550.7453 & 0.352 &  $-$88.07 & $-$0.94 & & 48483.6702 & 0.421 & $-$91.45  &  0.08 \\
45571.8844 & 0.396 &  $-$90.83 &  0.41 & & 48491.7518 & 0.437 & $-$93.09  &  1.02 \\
45602.8697 & 0.460 &  $-$94.12 &  1.06 & & 48517.7439 & 0.491 & $-$95.70  &  1.23 \\
45602.8747 & 0.460 &  $-$94.01 &  0.95 & & 48542.8674 & 0.543 & $-$96.94  &  0.10 \\
46507.4955 & 0.332 &  $-$89.19 &  0.69 & & 48607.4976 & 0.677 & $-$103.07 &  1.61 \\
46627.8390 & 0.581 &  $-$98.01 & $-$0.44 & & 48637.5401 & 0.739 & $-$103.18 &  0.98 \\
46663.7130 & 0.655 & $-$100.69 & $-$0.26 & & 48663.4661 & 0.793 & $-$102.35 &  0.39 \\
46682.7537 & 0.694 & $-$102.54 &  0.77 & & 48673.4993 & 0.813 & $-$101.55 & $-$0.10 \\
46712.5916 & 0.756 & $-$102.55 &  0.34 & & 48847.7009 & 0.174 & $-$87.50  & $-$0.88 \\
46739.7623 & 0.812 & $-$102.46 &  0.80 & & 48877.7406 & 0.236 & $-$87.13  & $-$0.45 \\
46779.5052 & 0.895 &  $-$99.46 &  0.06 & & 48908.5544 & 0.300 & $-$86.95  & $-$0.96 \\
46801.5491 & 0.940 &  $-$97.96 &  0.38 & & 48940.6623 & 0.366 & $-$89.66  &  0.24 \\
46832.4733 & 0.004 &  $-$95.54 &  0.84 & & 48965.6780 & 0.418 & $-$91.07  & $-$0.20 \\
46842.5687 & 0.025 &  $-$93.17 & $-$0.57 & & 48995.4879 & 0.480 & $-$94.40  &  0.46 \\
46957.8306 & 0.263 &  $-$87.34 & $-$0.24 & & 49024.4984 & 0.540 & $-$95.96  & $-$0.73 \\
47020.7667 & 0.394 &  $-$90.21 & $-$0.13 & & 49164.8304 & 0.830 & $-$102.65 &  1.34 \\
47070.6224 & 0.497 &  $-$94.35 & $-$0.39 & & 49230.7792 & 0.966 & $-$96.44  &  0.02 \\
47094.7307 & 0.547 &  $-$96.78 & $-$0.23 & & 49289.7675 & 0.088 & $-$92.27  &  1.26 \\
47128.5941 & 0.617 &  $-$98.87 & $-$0.93 & & 49317.6956 & 0.146 & $-$89.75  &  0.70 \\
47166.4648 & 0.695 & $-$101.10 & $-$0.69 & & 49349.6200 & 0.212 & $-$87.84  &  0.08 \\
47190.4773 & 0.745 & $-$102.61 &  0.40 & & 49372.5587 & 0.260 & $-$88.81  &  1.25 \\
47191.5694 & 0.747 & $-$102.45 &  0.23 & & 49507.8143 & 0.540 & $-$96.79  &  0.10 \\
47219.5727 & 0.805 & $-$101.34 & $-$0.44 & & 49537.8320 & 0.602 & $-$99.13  & $-$0.13 \\
47227.4895 & 0.821 & $-$101.50 &  0.01 & & 49550.8388 & 0.629 & $-$99.76  & $-$0.43 \\
47375.8293 & 0.128 &  $-$90.64 &  1.05 & & 49580.7970 & 0.691 & $-$101.14 & $-$0.57 \\
47390.8013 & 0.159 &  $-$88.56 & $-$0.15 & & 49611.7532 & 0.755 & $-$101.48 & $-$0.74 \\
47428.7699 & 0.238 &  $-$87.48 & $-$0.09 & & 49636.7966 & 0.806 & $-$101.79 &  0.03 \\
47459.7125 & 0.302 &  $-$87.53 & $-$0.41 & & 49700.6135 & 0.939 & $-$98.67  &  1.02 \\
47481.6151 & 0.347 &  $-$87.42 & $-$1.46 & & 49757.4569 & 0.056 & $-$93.10  &  0.75 \\
47511.5538 & 0.409 &  $-$90.95 &  0.03 & & 49903.8466 & 0.359 & $-$88.86  & $-$0.35 \\
47540.5805 & 0.469 &  $-$93.09 & $-$0.39 & & 49912.7672 & 0.377 & $-$88.82  & $-$0.96 \\
47567.4753 & 0.525 &  $-$94.03 & $-$2.00 & & 49920.7383 & 0.394 & $-$90.88  &  0.53 \\
47727.8202 & 0.857 & $-$100.36 & $-$0.27 & & 49938.8367 & 0.431 & $-$92.17  &  0.35 \\
47759.7682 & 0.923 &  $-$98.04 & $-$0.27 & & 49946.7754 & 0.448 & $-$90.55  & $-$1.98 \\
47775.8408 & 0.956 &  $-$97.52 &  0.63 & & 49964.8678 & 0.485 & $-$94.49  &  0.28 \\
47804.7104 & 0.016 &  $-$94.51 &  0.35 & & 49973.7108 & 0.504 & $-$93.88  & $-$1.17 \\
47832.5893 & 0.073 &  $-$92.02 &  0.40 & & 49980.6262 & 0.518 & $-$95.43  & $-$0.28 \\
47864.5926 & 0.140 &  $-$88.95 & $-$0.29 & & 50000.7224 & 0.559 & $-$97.82  &  0.26 \\
47895.5659 & 0.204 &  $-$87.89 &  0.03 & & 50006.7072 & 0.572 & $-$98.26  &  0.18 \\
47923.4492 & 0.261 &  $-$87.44 & $-$0.13 & & 50026.6268 & 0.613 & $-$99.43  & $-$0.24 \\
48056.8368 & 0.537 &  $-$97.47 &  0.88 & & 50054.6611 & 0.671 & $-$101.47 &  0.14 \\
48083.8350 & 0.593 &  $-$99.72 &  0.78 & & 50097.4519 & 0.760 & $-$101.06 & $-$1.15 \\
48108.8078 & 0.645 & $-$101.41 &  0.73 & & 50111.5168 & 0.789 & $-$101.20 & $-$0.80 \\
48131.8045 & 0.693 & $-$101.85 &  0.11 & & 50139.4779 & 0.847 & $-$100.42 & $-$0.49 \\
48158.7417 & 0.748 & $-$101.73 & $-$0.49 & & 50293.7783 & 0.166 & $-$87.83  & $-$0.72 \\
48163.7844 & 0.759 & $-$101.33 & $-$0.88 & & 50325.6689 & 0.232 & $-$86.96  & $-$0.64 \\
48191.6865 & 0.816 & $-$101.05 & $-$0.54 & & 50363.7428 & 0.311 & $-$89.05  &  0.97 \\
48221.5286 & 0.878 &  $-$99.29 & $-$0.68 & & 50374.7099 & 0.333 & $-$89.78  &  1.25 \\
48257.6683 & 0.953 &  $-$96.04 & $-$0.98 & & 50405.6227 & 0.397 & $-$91.51  &  1.04 \\
48284.4812 & 0.008 &  $-$93.81 & $-$0.69 & & 50438.5882 & 0.465 & $-$94.69  &  1.39 \\
\enddata
\tablecomments{MJD = JD - 2400000, where JD is the Julian Date. 
The units for $v_r$ and O--C are \kms.
The orbital phase is the fractional part of 
$\phi = ({\rm JD} - T_0) / P$, where JD is the 
Julian Date listed above,
$T_0$ = JD 2450213.508, and $P$ = 483.3~d.}
\label{tab: rvs}
\end{deluxetable}
\clearpage

\begin{deluxetable}{lcccccc}
\tablecolumns{7}
\tablecaption{EG And Orbital Solutions}
\tabletypesize{\footnotesize}
\tablehead{{Data Set} & {$P_{orb}$ (d)} & {$\gamma$ (\kms)} & 
{$K$ (\kms)} & {$T_0$} & {$a$~sin~$i$ (AU)} & {$f(M)$ (\msun)} }
\startdata
All &   482.5 $\pm$ 1.3 & $-$94.74 $\pm$ 0.09 & 7.30 $\pm$ 0.13 & 
208.108 $\pm$ 0.672 & 0.323 $\pm$ 0.003 & 0.019 $\pm$ 0.001 \\
Fekel & ~483.2 $\pm$ 13.3 & $-$95.18 $\pm$ 0.12 & 7.66 $\pm$ 0.22 & 
198.797 $\pm$ 4.160 & 0.340 $\pm$ 0.013 & 0.023 $\pm$ 0.003 \\
SAO &   483.3 $\pm$ 1.6 & $-$94.88 $\pm$ 0.05 & 7.34 $\pm$ 0.07 &
213.508 $\pm$ 0.961 & 0.326 $\pm$ 0.003 & 0.020 $\pm$ 0.001 \\
OMMF &   482.7 $\pm$ 1.3 & $-$95.01 $\pm$ 0.10 & 7.28 $\pm$ 0.18 &
200.768 $\pm$ 1.840 & 0.323 $\pm$ 0.008 & 0.019 $\pm$ 0.001 \\
\enddata
\tablecomments{
The time of spectroscopic conjunction is 2450000 + $T_0$.
Data sets are
`All': all published radial velocities, including those in this paper;
`Fekel': KPNO data from \citet{fekel2000};
`SAO': data from this paper; and
`OMMF': data from \citet{oliver1985}, \citet{munari1988}, 
\citet{munari1993}, and \citet{fekel2000}.
}
\label{tab: orbs}
\end{deluxetable}
\clearpage

%
%\centerline{\bf FIGURES}
%

\clearpage

\clearpage

\clearpage

\clearpage

\clearpage

\clearpage

\end{document}